\def\avrg#1{{\langle{#1}\rangle}}

\documentstyle[12pt,aasms4]{article}

\begin{document}

\title{A Robust Classification of Galaxy Spectra: Dealing with 
Noisy and Incomplete Data}

\author{A.J. Connolly\altaffilmark{1} and A.S. Szalay}
\affil{Department of Physics and Astronomy, The Johns Hopkins
University, Baltimore, MD 21218}
\authoremail{ajc@pha.jhu.edu, szalay@jhu.edu}
\altaffiltext{1}{Department of Physics and Astronomy, 
University of Pittsburgh, Pittsburgh, PA, 15260}

\begin{abstract}
Over the next few years new spectroscopic surveys (from the optical
surveys of the Sloan Digital Sky Survey and the 2 degree Field survey
through to space-based ultraviolet satellites such as GALEX) will
provide the opportunity and challenge of understanding how galaxies of
different spectral type evolve with redshift. Techniques have been
developed to classify galaxies based on their continuum and line
spectra.  Some of the most promising of these have used the Karhunen
and Lo\'{e}ve transform (or Principal Component Analysis) to separate
galaxies into distinct classes. Their limitation has been that they
assume that the spectral coverage and quality of the spectra are
constant for all galaxies within a given sample. In this paper we
develop a general formalism that accounts for the missing data within
the observed spectra (such as the removal of sky lines or the effect
of sampling different intrinsic rest wavelength ranges due to the
redshift of a galaxy). We demonstrate that by correcting for these
gaps we can recover an almost redshift independent classification
scheme. From this classification we can derive an optimal
interpolation that reconstructs the underlying galaxy spectral energy
distributions in the regions of missing data.  This provides a simple
and effective mechanism for building galaxy spectral energy
distributions directly from data that may be noisy, incomplete or
drawn from a number of different sources.
\end{abstract}


\section{Introduction}

The next generation of spectroscopic surveys, such as the Sloan
Digital Sky Survey (SDSS) and the 2 degree Field redshift survey (2dF;
Maddox et al.\ 1998) will provide a wealth of information about the
spectral properties of galaxies in the local and intermediate redshift
universe. For the first time we will have high signal-to-noise
spectrophotometry of large, systematically selected samples of
galaxies. The quality of these data will be such that we will not be
restricted to measuring just redshifts but will be able to extract the
spectral characteristics of individual galaxies.  If we can define
robust methods for the classification of galaxy spectra we will be
able to study the evolution of the spectral properties of galaxies and
relate these observations to the physical processes that drive them.

In the light of this a number of statistical techniques have been
developed for automated classification of galaxies based on spectral
continuum and line properties. One of the most promising of these
methods has been the Principal Component Analysis or
Karhunen-Lo\`{e}ve transform (Karhunen 1947, Lo\`{e}ve 1948). The
technique has been successfully applied to the classification of
galaxy spectra (Connolly et al. 1995, Folkes et al.\ 1996, Sodre and
Cuevas 1997, Bromley et al.\ 1998, Galaz and de Lapparent, 1998), QSOs
(Francis et al. 1992) and stars (Singh et al.\ 1998).

The underlying basis behind these techniques is that a galaxy spectrum
can be described by a small number of orthogonal components
(eigenspectra). These eigenspectra are found to correlate strongly
with the physical properties of galaxies, such as the star formation
rate or age of the stellar population. By projecting a galaxy spectrum
onto these orthogonal components we have a measure of the relative
contributions of these different stellar types and consequently an
estimate for the spectral type of the galaxy.  Since the spectral
energy distribution of galaxies is the sum of the SEDs of their
stellar population, such an approach is quite natural and should give
a reasonable description of the galaxies.

Each of these approaches makes the underlying assumption that the
galaxy spectra are perfect. In the real world, where surveys will
cover a wide range in redshift and luminosity, this will not be the
case. The ensemble of galaxy spectra will cover a broad range of rest
wavelengths, have variable signal-to-noise, and will contain spectral
regions affected by sky lines or artifacts in the spectrographs. This
will result in spectra whose wavelength coverage will only be a subset
of that of the eigenspectra (i.e.\ the data will contain missing
spectral regions or gaps within the spectra). Applying the standard
techniques whereby we project the galaxy spectra onto an eigenbasis
with a simple scalar product will introduce biases in to the galaxy
classification schemes.

In this paper we address these issues. We extend the KL analysis of
spectra to incorporate the effects of gappy data and variable signal
to noise. In section 2 we provide the mathematical basis for our
analysis and in sections 3 and 4 we show how we can provide an optimal
interpolation of galaxy spectra over the missing data. Sections 5 and
6 demonstrate how eigenspectra can be built from noisy and incomplete
spectra. Finally in Section 7 we discuss the application of these
techniques to the general case of incomplete data and how they might
be used in astrophysical problems.

\section{An Orthogonal Expansion of Censored Data}

It has been known for some time that galaxy spectra can be represented
by a linear combination of orthogonal spectra (i.e.\ eigenspectra) and
that these eigenspectra can be used for galaxy classification.  In an
earlier paper (Connolly et al.\ 1995) we described the technique for
applying the Karhunen-Lo\`{e}ve transform (KL; also known as Principal
Component Analysis) to derive these eigenspectra. In this paper we
will discuss the application of the KL transform to the classification
of galaxies when we have gaps in the spectra (e.g.\ due to the removal
of sky lines, bad regions on a spectrograph's CCD camera or galaxies
with different rest wavelength coverage).

From the KL transform we can construct an orthonormal eigenbasis such
that each galaxy spectrum, $f_\lambda$, can be represented as a linear
combination of eigenspectra, $e_{i\lambda}$.
\begin{equation}
	f_\lambda = \sum_i a_i e_{i\lambda},
\end{equation}
where $\lambda$ represents the wavelength dimension, $i$ is the number
of the eigenspectrum and $a_i$ are the coefficients of the linear
combination.  If we project a galaxy onto this eigenbasis we can
define the set of linear expansion coefficients, $a_i$, that fully
describe a galaxy spectrum. The eigenspectra $e_{i\lambda}$ are
defined to be orthogonal such that, using a simple scalar product,
\begin{equation}
	\sum_\lambda e_{i\lambda} e_{j\lambda} = \delta_{ij},
\end{equation}
where the sum in $\lambda$ extends over a pre-defined wavelength
range $(\lambda_1,\lambda_2)$. The eigenspectra are ranked by
decreasing eigenvalues, which in turn reflect the statistical significance
of the particular eigenspectrum. For the details of the construction
of the eigenbasis see Connolly et al.\ (1995).

We have shown that, for galaxy spectra, the majority of the
information present within the data is contained within the first 3-7
eigencomponents (Connolly et al.\ 1995). Consequently the expansion of
each galaxy spectrum in terms of the eigenbasis can be truncated
(i.e.\ we can retain most of the information present within the galaxy
data using only a handful of components). This truncated expansion
represents an optimal filtering of the data in the least squares
sense. As such it provides a very efficient mechanism for describing
galaxy spectra.

From this series of expansion coefficients, whether truncated or not,
we can construct a very natural classification scheme. As we have
shown (Connolly et al.\ 1995), the first three coefficients correlate
with the amount of a galaxy spectral energy distribution that is
dominated by an old stellar population or by active star
formation. These coefficients form an approximately one parameter
sequence, well correlated with the ages of galaxies and the
distribution of these coefficients can be used to separate galaxies
into distinct classes (e.g. Folkes et al.\ 1996, Bromley et al.\
1998).

The underlying basis behind these analyses is that a galaxy can be
represented as a linear combination of orthogonal spectral components.
The orthogonality of the system is important as it means that the
expansion coefficients are uncorrelated and, therefore, provide a very
simple and general way to separate galaxies into distinct classes. The
benefits of the orthogonal expansion only hold if the eigenspectra and
the galaxy spectrum are constructed over the same wavelength
range. If, as found in real spectra, there are regions of missing data
within a galaxy spectrum, due, for example, to the presence of sky
lines or the fact that the rest wavelength coverage of the galaxy
spectra differ from that of the eigenspectra, then the orthogonality
of the system no longer holds. This loss of orthogonality can be
understood if we consider Figure 1. The eigenbasis (as shown by the
first 3 eigenspectra) is clearly orthogonal over the full spectral
range (the scalar product between the individual eigenspectra is
zero). If we exclude those data points with $\lambda < 5000$ \AA\ we
see that the orthogonality no longer holds and the eigensystem is not
linearly independent. This means that the coefficients that would be
derived by simply projecting a galaxy spectrum onto this censored
basis would be biased (in other words the spectral modes become
correlated). As the wavelength range over which the data can be
defined as valid clearly varies as a function of redshift and
spectrograph, comparison between the classification of different
galaxy populations becomes extremely difficult.

For the case of reconstructing faces from gappy data Everson and
Sirovich (1995) have shown that we can account for the gaps within the
data and reconstruct ``unbiased'' correlation coefficients. We extend
here the analysis of Everson and Sirovich (1995) and Connolly and
Szalay (1996) to the case of spectral data and generalize the problem
to consider an arbitrary weighting of the spectra.

We consider the gappy galaxy spectrum, $f^o_\lambda$, that we wish to
project onto the eigenbasis, as consisting of the true spectrum (i.e.\
without gaps), $f_\lambda$, and a wavelength dependent weight,
$w_\lambda$. The wavelength dependent weight will be zero where there
are gaps within the data (corresponding to infinite noise), and
$1/\sigma_\lambda^2$ for the remaining spectral range. By applying
this weight function we have a general mechanism by which we can down-
or up-weight not just bad regions but also particular spectral
features (e.g.\ emission lines) that we wish to emphasize within the
data. It is worth noting that in the special case where we do not
consider the effects of noise (i.e.\ the weight values are 0 or 1)
then the wavelength dependent weight acts as a window function and
$f^o_\lambda = w_\lambda f_\lambda$. In the general case that we
discuss below, where we include gaps and noise, the weight function is
related to the true spectrum through the $\chi^2$ minimization
(Equation 3).

Given the relative weight of each spectral element we wish to derive a
set of expansion coefficients that minimize the quadratic deviation
between the observed spectrum, $f^o_\lambda$, and its truncated
reconstruction, $\sum_i a_i e_{i\lambda}$, where the sum over $i$
extends to a small number of eigenspectra only. To do this we define
the $\chi^2$ statistic such that,
\begin{equation}
	\chi^2 = \sum_\lambda w_\lambda 
	(f^o_\lambda - \sum_i a_i e_{i\lambda})^2
\end{equation}
and minimize this function with respect to the $a_i$'s.  This gives
the minimal error in the reconstructed spectrum, over the full range in
$\lambda$, weighted by the variance vector, $w_\lambda$.

Solving for $a_i$ we get,
\begin{equation}
	\sum_\lambda w_\lambda e_{j\lambda} \sum_i a_i  e_{i\lambda} = 
	\sum_\lambda w_\lambda f^o_\lambda e_{j\lambda},
\end{equation}
Defining $M_{ij} = \sum_\lambda w_\lambda e_{i\lambda} e_{j\lambda}$
and $F_j = \sum_\lambda w_\lambda f^o_\lambda e_{j\lambda}$ this
simplifies to
\begin{equation}
	a_i = \sum_j M_{ij}^{-1} F_{j}.
\end{equation}
 
Clearly $F_j$ represents the expansion coefficients that we would have
derived if we had undertaken a weighted projection of the observed
galaxy spectrum $f^o_\lambda$ onto the eigenbasis (i.e.\ a biased set
of coefficients) and $M_{ij}^{-1}$ tells us how the eigenbasis is
correlated over the censored spectral range. If the weights were all
equal (there was no region that was masked or of lower
signal-to-noise) then Equation 4 simplifies to that given in our
original Equation 1 and $M_{ij}^{-1}$ becomes a unity matrix. As we
introduce gaps into the spectra the off-diagonal components of
$M_{ij}^{-1}$ become more significant.

Therefore, by correcting for the correlated nature of the eigenbasis
we can determine the values of the expansion coefficients, $a_i$ that
we would have derived had we had complete spectral coverage and no
noise within the observed spectra. As such they are independent
(within the errors) of the wavelength range over which we observe a
galaxy and can be used to classify galaxy spectra taken over a wide
range in redshift and with differing signal-to-noise. This enables an
objective comparison of galaxy spectral types using the complete
spectral information and free from the wavelength dependent selection
biases that may be present in existing analyses.

Associated with the corrected expansion coefficients $a_i$ we can
define a covariance matrix. This measures the uncertainty in the
coefficients due to the correlated nature of the eigensystem. It is
straightforward to show that the covariance between the expansion
coefficients is simply,
\begin{equation}
	{\rm Covar}(a_i a_j) = \avrg{a_i a_j} - \avrg{a_i}\avrg{a_j}
              = \frac{1}{N} M^{-1}_{ij}
\end{equation}
where $N=\sum_\lambda 1/\sigma_\lambda^2$. The size of the uncertainty
in the expansion coefficients, after the correction, is proportional
to the amount that the eigenbasis is correlated (as we would
expect). From this analysis we, therefore, have a correction for the
effect of the gaps inherent within real spectra and a measure of the
error on these derived values.

Given this approach we can not only derive the set of corrected
coefficients for the classification of galaxy spectra, but we can also
use these coefficients to reconstruct the regions of the spectra that
are masked. This tells us that if a galaxy spectrum can be reproduced
using a handful of components then the spectral features present
within the data (e.g.\ the Balmer series of absorption lines) are
correlated (as we would expect from the physics). Therefore, if we
have sufficient spectral coverage to detect a feature in the spectrum
we can predict the strengths of additional features where we have no
data. The gappy KL analysis does this in a mathematically rigorous
way, allowing the data themselves to define the inherent correlations.

\section{Optimal Interpolation of Gaps in the Spectra}

Section 2 outlines the basic mathematical and physical reasoning
behind the classification of galaxy spectra in the case of gappy data.
We now consider the application of these techniques to spectral data.
In order to be able to test our technique, we create an eigenbasis,
using the GISSEL96 model spectral energy distributions of Bruzual and
Charlot (1993). We use a simple stellar population model with an
instantaneous burst of star formation at zero age and sample the model
spectra from 0 to 20 Gyr. In total the sample contains 222 spectra
covering the wavelength range 3500 \AA\ to 8000 \AA\ (designed to
approximate the spectral coverage of the SDSS data). The choice of our
particular Bruzual and Charlot model is somewhat arbitrary as we are
only concerned in having a set of spectra that cover a wide range in
age and for which we can control the uncertainties within the data.

We construct the eigenbasis as described in Connolly et al.\ (1995)
for the Bruzual and Charlot data after normalizing all spectra to unit
scalar product. The diagonalization of the correlation matrix is
undertaken using the Singular Value Decomposition algorithms of the
Meschach package. In Figure 1 we show the first three eigenspectra and
in Figure 2 the corresponding sequence of eigenvalues. The size of the
eigenvalue is directly related to the amount of variance (or
information) contained within each of the eigenspectra. The
eigenvalues decrease rapidly with the first three components
containing 99.97\% of the total system variance.  By the tenth
eigenspectrum the eigenvalue (or variance of the system contained
within this spectrum) is at the level of $10^{-4}$ of a percent.
Using just the first three eigencomponents (i.e.\ truncating the
expansion) we should, therefore, be able to reconstruct any given
spectrum to an accuracy of better than 0.05\%.

Considering these eigenspectra in turn, the first eigenspectrum is the
mean spectrum and represents the `center of mass' of the Bruzual and
Charlot sample of galaxy spectra. The second eigenspectrum has the
spectral shape of an O star and describes the star formation component
of the galaxy spectral energy distribution. The third component is a
mixture of an old G or K star stellar population (with a strong 4000
\AA\ break) and an intermediate age A star population (with strong
Balmer lines). From the distribution of eigenvalues (see above) a
linear combination of these three stellar spectra can, therefore,
reproduce the full range of the Bruzual and Charlot spectral energy
distributions to a very high accuracy.

If we project a galaxy onto this eigenbasis the expansion coefficients
tell us the relative contributions of each of these components. This
provides not only a classification of the galaxy but also a means of
reconstructing the underlying spectrum.  As we can reproduce the
galaxy spectra with a small number of eigenspectra we should be able
to use these components to interpolate over regions without data.

For the case of real spectra we might expect the KL reconstruction to
require more than just the handful of eigencomponents that we need for
the synthetic data (e.g.\ to account for the distinct spectral
signatures of the small number of AGN present within any spectroscopic
survey). The techniques we will apply here should be equally
applicable to real spectra given the provisos that the eigenbasis will
provide a better reconstruction if it is built from similar types of
galaxies and that the number of components required may be
significantly larger (with the associated increase in computational
resources).

\subsection{Interpolation due to missing data}

As we have shown in Section 2 a simple projection of a galaxy spectrum
onto the eigenbasis will result in a biased set of expansion
coefficients. If, however, we account for the gaps within the data, we
can correct the expansion coefficients and use these values to
estimate the underlying spectrum. In the following analysis we will
consider the case of randomly positioned gaps within a galaxy
spectrum. This is designed to simulate the effect of excluding
spectral regions due to the position of sky lines or artifacts in a
spectrograph.  We initially assume that we know the underlying
eigenbasis that describes the galaxy populations (in section 6 we will
expand this analysis to construct the eigenbasis itself from gappy
data) and we ignore the effect of noise.

In Figure 3 we take three representative spectra, a zero age spectrum
an intermediate age spectrum (0.16 Gyr) and an old stellar population
(20 Gyr). Each of these spectra has been drawn from the sample of 222
galaxies in the Bruzual and Charlot sample described above. For each of
these three spectra we mask the wavelength range 3800 \AA\ to 4000
\AA. We project each spectrum onto the eigenbasis over the spectral
range 3500 \AA\ to 8000 \AA\ (excluding the masked region). We then
correct these derived coefficients for the correlated nature of the
eigenbasis (i.e.\ due to the masked regions). Figure 3 shows the
reconstruction of the galaxy spectra within the masked region, when
using 3 and 5 eigencomponents respectively. The solid line shows the
true spectrum and the dotted line the reconstruction.  To compare the
accuracy of the reconstruction as a function of galaxy type we define
an error that is independent of the overall galaxy flux. This error is
given by the rms deviation between the reconstructed and ``true''
spectra (over the masked region) when both spectra are normalized by
their scalar product to unity.

For three eigencomponents the reconstruction works well for the 0 Gyr
and 0.16 Gyr galaxy spectra. The normalized rms deviation between the
true spectrum and the reconstructed data is only 0.0016. For the 20
Gyr model the reconstructed spectrum produces the correct features
present within the data (i.e.\ the absorption lines present in the
true spectrum are found in the reconstruction) but their relative
amplitudes are inconsistent. Over the masked spectral range the
deviation between the reconstructed and true spectrum for this 20 Gyr
model is 0.018, a factor of ten worse than the younger stellar
types. Given that we are deriving the interpolation based on three
spectral components that are constructed over the full wavelength
range 3500 \AA\ to 8000 \AA\ it is remarkable that we can reproduce
the observed spectra to such high accuracy.

If we incorporate a further two eigencomponents ($N_{eigen} = 5$) the
reconstruction of the 20 Gyr model is substantially improved. The
deviation between the true and reconstructed spectrum falls to only
0.0045. The other two spectral types also show improvement in the
interpolation though the magnitude of this improvement is not as
dramatic as for that of the 20 Gyr spectrum. The first three
eigenspectra are, therefore, more sensitive to the spectral features
(both continuum and absorption lines) of star forming and intermediate
age galaxies. This is not entirely surprising as the Bruzual and
Charlot models from which we construct the eigenbasis are dominated by
these types of galaxies (over 135 of the 222 spectra come from
galaxies with ages less than 1 Gyr). The eigenspectra will, therefore,
be weighted more towards these younger galaxy spectra than to the more
evolved stellar populations.

\subsection{Interpolation due to the effect of redshift}

While the interpolation of galaxy spectra across narrow spectral
intervals may be seen as relatively straightforward a more challenging
problem is how to extrapolate a galaxy spectrum. The need for this
will arise if we wish to project a galaxy spectrum onto an eigenbasis
where the galaxy's wavelength coverage is only a subset of that of the
eigenspectrum (e.g.\ the eigenspectra and galaxy are at different
redshifts).

In Figure 4 we demonstrate that the correction for gappy data can
equally apply to the case of extrapolating a galaxy spectrum as well
as for simple interpolation. For the 0 Gyr, 0.16 Gyr and 20 Gyr galaxy
spectra we exclude the wavelength range 7100 \AA\ to 8000 \AA\ and
apply the KL reconstruction as described above. As before, the solid
line shows the true spectrum and the dotted line the reconstructed
spectrum. The left hand panel shows the reconstruction when using
three eigencomponents and the right hand panel for five components.

The 0 Gyr, 0.16 Gyr and 20 Gyr spectra are all accurately
reconstructed from three components. The rms uncertainty in the 0 Gyr
0.16 Gyr and 20 Gyr spectra amount to 0.0016, 0.0006 and 0.001
respectively. The 0 Gyr and 20 Gyr models are systematically offset by
approximately 4\% when using three eigencomponents. As found in
Section 3.1 increasing the number of eigenspectra used in the
reconstruction improves the resulting spectra for all three galaxy
types. The most marked improvement occurs for the 0 Gyr and 20 Gyr
models where the deviation drops to less than 0.0002.

The results are, unsurprisingly, similar to those found for the case
of simple interpolation. All three spectral types are well described
by 3 eigencomponents. Of these the 0 Gyr and 20 Gyr spectra have the
largest errors. Increasing the number of components used in the
reconstruction reduces the dispersion between the true and corrected
spectrum decreases. It is, therefore, clear that projecting a galaxy
spectrum onto its eigenbasis gives a natural (and optimal in the
quadratic sense) interpolation scheme. It utilizes the correlations
inherent within the data to determine how individual spectral regions
are related.

\section{Galaxy Classification using Gappy Data}

Projecting a galaxy onto its eigenbasis provides a very simple and
natural classification scheme. As we have shown in Section 3 the
eigenspectra are highly correlated with stellar spectral types (with
the second and third eigenspectra correlating with O and K stellar
spectral energy distributions). By projecting a galaxy spectrum onto
this eigenbasis the expansion coefficients, $a_i$, will tell us the
contribution of each of these eigenspectra to the overall spectral
energy distribution. As has been shown previously, these expansion
coefficients can then be used to separate galaxies into distinct
spectral classes (Connolly et al.\ 1995, Folkes et al., 1996, Bromley
et al., 1998).

In Figure 5a we demonstrate this effect for the Bruzual and Charlot
model. The solid line shows the distribution of the first two
expansion coefficients, $a_1$ and $a_2$ as a function of galaxy
age. The galaxies form a simple, one parameter distribution, that
transitions from star forming galaxies (bottom of the plot) to
quiescent, old stellar populations (top of the plot).  For the simple
stellar population given by the Bruzual and Charlot model the
correlation between expansion coefficients and galaxy age is extremely
tight. In the case of real data we find that there is a much larger
dispersion in the relation (Bromley et al., 1998).

Some of this dispersion may be due to the failure to correct for the
gappy nature of real spectra. As we have shown in section 2, when
galaxy spectra are projected onto an eigenbasis without correcting for
the gaps within the data the eigenbasis is no longer orthogonal. This
means that the eigenspectra become correlated. The consequence of this
is that the expansion coefficients are also correlated and any
classification scheme based on gappy data will be biased. As we will
show, the biasing of the expansion coefficients due to the gappy
nature of the data does not just introduce a larger statistical
uncertainty into any derived classification, it can also produce
systematic errors.

We initially consider the case of spectra with missing spectral
regions due to the presence of sky lines or defects in the
spectrograph (as in Section 3.1). The gaps within the data will
manifest as small wavelength regions where the galaxy spectrum must be
masked or interpolated across. We simulate this using the Bruzual and
Charlot data by excising ten randomly positioned spectral regions,
each of 45 \AA\ in width. The effect of this masking on the derived
expansion coefficients is shown in Figure 5a. The coefficients
derived from the masked data are shown by crosses.

Masking of these spectral regions introduces a significant dispersion
into the classification scheme. For the 10\% masking adopted above,
the dispersion in the classification correlation is approximately 0.03
in absolute number or a 3\% error in terms of the sum of the square of
the coefficients (the coefficients sum, in quadrature, to unity due to
the scalar product normalization). If we apply the corrections to the
expansion coefficients as described in section 2 we can reconstruct
the original unbiased coefficients. In Figure 5a the corrected
coefficients are shown by the filled ellipses. The size of the
ellipses are defined by the three sigma errors on the corrected
expansion coefficients as calculated from the variance analysis in
Equation 6. We find that by applying the corrections to the
coefficients we can recover the underlying true expansion coefficients
and thereby derive an unbiased classification.

A more important effect in terms of the classification of galaxies is
the effect of redshift. When we analyze an ensemble of galaxies over a
range of redshifts, the intrinsic rest wavelengths that we sample
will be dependent on the redshift of the galaxy in question. In
principle we could just consider those spectral regions that are in
common to all galaxies within our sample. In practice, however, the
wide range in redshifts that we will be faced with in the 2dF and SDSS
surveys may result in very little wavelength overlap for galaxies at
the extremes of the redshift distribution (e.g.\ the redshift
distribution for the SDSS is expected to have a significant tail of
$z>0.5$ galaxies which would reduce the wavelength range common to all
galaxies by $\sim$40\%).

In Figure 5b we simulate the effect of redshift on the classification
of galaxies. We assume that the galaxies are randomly distributed
between redshifts of $0<z<0.2$ (a conservative assumption) and mask
out those regions of the spectrum that lie beyond the 8000 \AA\ cutoff
(see section 3). The effect on the derived coefficients due to these
censored data is shown by the crosses in Figure 5b. In contrast to the
effect of randomly positioned gaps within a galaxy spectrum (which
introduce a random scatter into the classification coefficients) we
find that the effect of redshift is to systematically bias the
expansion coefficients. The first expansion coefficient, $a_1$, is
systematically over estimated by approximately 10\% and the second
component underestimated by approximately 25\%. As before, the solid
ellipses in Figure 5b show the expansion coefficients once corrected
for the missing spectral regions. The size of the points reflect the
three sigma uncertainties in the corrected coefficients due to the
variance determined from Equation 6.

Therefore, if we apply the correction for the gaps within the observed
data we can reproduce, to high accuracy, the underlying classification
coefficients. Given our current, conservative, simulation where we
excise up to 1500 \AA\ we can recover the true coefficients to an
accuracy of better than 0.002 in absolute number or 0.2\%. Increasing
the amount of data that we mask will naturally make the
eigenspectra more correlated and the derived coefficients less
accurate. 

\section{Reconstructing Spectral Energy Distributions from Noisy Data}

In the previous sections we assumed that the observed galaxy spectrum
was gappy but free of noise (essentially assuming that the weight
function was zero or unity). The general form of Equation 2 enables us
to extend these analyses to account for galaxy spectra in the presence
of noise. We demonstrate here that the use of the KL expansion
provides an optimal filtering of a noisy spectrum and that the
correction for gaps within a spectrum is equally applicable in the
presence of noise.

In the top panel of Figure 6 we show a 20 Gyr spectrum with a
signal-to-noise of approximately 5 per pixel (the noise is constant as
a function of wavelength). We project the spectrum onto the first
three eigenspectra (as derived in Section 3) and determine the
expansion coefficients. From this truncated expansion we can
reconstruct the underlying galaxy spectral energy distribution. The
reconstructed spectrum is shown in the lower panel of Figure 6 (dotted
line) together with the true, noise free, spectrum (solid line). A
comparison of the true and reconstructed spectra shows that they are
consistent with a total deviation of 0.00039. Even given significant
amounts of noise (on a pixel level) we can, therefore, reconstruct the
underlying spectral energy distribution with a high level of accuracy.

The reason the reconstruction reproduces the galaxy spectrum so
accurately is straightforward to understand if we consider the
integrated signal-to-noise of the full spectrum. If each galaxy can be
reproduced by 3-5 eigencomponents then we can describe a galaxy
spectrum by at most 5 numbers. In principle we should, therefore, only
require 5 data points on a spectrum to constrain these eigencomponents
(in practice with only a small number of data points the eigenspectra
become very correlated and the uncertainty in the derived expansion is
large). Even in the case of substantial noise (per pixel) we can
co-add the pixels to reduce the overall noise on the expansion.
Applying this truncated expansion to real life observations provides
an optimal (in the least-squares sense) filtering of the data and
should provide a substantial improvement in signal-to-noise.

For the case of gaps within noisy spectra we can still reconstruct
missing spectral regions. In Figure 7 we reproduce the analysis of
Section 3.1 for a 20 Gyr spectrum. We exclude those data in the
wavelength range 3800 \AA\ to 4000 \AA\ and reduce the input spectrum
to a signal-to-noise of 5 (per pixel). The noisy spectrum is then
projected onto the first three eigenspectra and the expansion
coefficients corrected for the correlated eigenbasis.  Using the first
three eigenspectra we reconstruct the overall galaxy spectrum. The
reconstructed spectrum is shown in Figure 7 by a dotted line and the
true spectrum by a solid line. The reconstruction is almost identical
to that derived from the noise-free spectrum. In the region 3800 \AA\
to 4000 \AA\ the deviation between the reconstructed spectrum using
noisy data and the error free data is 0.019 (comparable to the noise
free case). 

Combining the optimal filtering of the KL truncated expansion with the
correction for gaps within the data we can, therefore, reproduce
galaxy spectra at very high signal-to-noise. A natural application for
this procedure is the filtering of low signal-to-noise data derived
from spectroscopic redshift surveys. Using the eigenbasis as a
template for cross-correlation with the observed galaxy spectrum
(Glazebrook et al.\ 1998) and correcting for the bias in the
classification coefficients we can derive an optimal representation of
the underlying spectrum, an estimate of the significance of the
correlation and a measure of the classification coefficients (which
describe how closely the galaxy is related to the overall distribution
of galaxy spectral types). This latter information is a quality
assurance test. If the classification coefficients do not lie within
the general distribution then the galaxy in question has either an
unusual spectral type (worth further study) or there is a miss-match
in the classification (and the redshift is probably incorrect).

How well we will be able to classify and repair galaxy spectra will
ultimately be limited by how we construct the eigenbasis itself.  If
the type of galaxies that we construct the eigenspectra from do not
fully sample the population of galaxies to which we wish to apply the
classification then there could be a significant mismatch between
the spectral properties of the eigentemplates and the galaxies being
reconstructed (e.g.\ if we tried to use normal galaxy spectra to
classify QSOs and AGNs then the reconstruction would be poor).  This
problem can, however, be overcome by building the eigenbasis using a
subset of galaxies that are selected to evenly sample the distribution
of galaxy types rather than being weighted by the relative strengths
of the different galaxy populations.  Even without this approach the
residuals between the observed and reconstructed spectra (within those
spectral regions that contain valid data) will provide a measure of
how well the eigenbasis can describe a particular galaxy spectrum and,
consequently, whether the classification is valid.

\section{Building Empirical Spectral Energy Distributions}

By correcting for the gaps within galaxy spectra we have derived a
simple mechanism for classifying galaxies (and interpolating across
the regions of missing data) that can be applied to spectra that do
not fully overlap. We can now extend the analysis to constructing an
eigenbasis from gappy data. The earlier derivation assumed that we
knew what the underlying eigenbasis should be. We also assumed that
the eigenspectra are well constrained over the full spectral range
that the galaxies cover.  This is a feasible proposition if we use
spectral synthesis models to derive a set of eigenspectra and then
project observed galaxy spectra onto this basis. This has the
advantage that one can relate the coefficients directly to the
physical properties of the models (e.g.\ age or metalicity). Its
disadvantage is that we know that the spectral synthesis models cannot
yet reproduce the observed colors of galaxies (particularly at high
redshift) and so may not describe fully the spectral properties of all
galaxy populations. Secondly, the models are generally derived from
intermediate resolution spectra (with a dispersion of $\sim$10 \AA)
while the new generation of spectroscopic surveys will have a
substantially higher resolution (e.g.\ 3 \AA\ for the
SDSS). Therefore, by restricting ourselves to model spectra we may
miss important physical information present within the spectral data.

In an ideal case we would want to build the set of eigenspectra
directly from the observed spectra. In such a way the data for the
eigenbasis and galaxies that we wish to classify would be taken
through the same optical system and have the same intrinsic
resolution. Unfortunately, relying on observations means that we must
construct a set of eigenspectra from data that have different
restframe spectral coverage (due to the redshift of the galaxies) and
gaps within the spectra (e.g.\ from the removal of sky lines). If the
galaxies occupy a range of redshifts then missing spectral regions
will occur at different rest wavelengths. Therefore, for a large
ensemble of galaxies over a range of redshifts, all spectral regions
will be sampled by a number of galaxies and the eigenspectra can be
constructed over the superset of wavelengths that the galaxies cover.

To build the eigenbasis we take an iterative approach in a manner
analogous to that described in Everson and Sirovich (1995). After
shifting all galaxy spectra to their restframe we linearly interpolate
across all gaps within the spectra. This gives the zeroth order
correction for each spectrum, $f_i^0$. From these corrected data we
build the eigenspectra, $e_i^0$, and project each of the individual
spectra onto this basis. After correcting the expansion coefficients
for the gappy nature of the projection we can use the KL basis $e_i^0$
to interpolate across the regions of missing data and form a first
order corrected spectrum, $f_i^1$. This procedure continues until
convergence.

The iterative technique used to construct the eigenbasis and then
repair the galaxy spectra is shown in Figure 8.  Each galaxy used to
create the eigenbasis has had 10 gaps of 100 \AA\ randomly positioned
within the spectrum. From this gappy data set we undertake the
iterative procedure outline above.  Figure 8 shows the spectrum of a 2
Gyr old galaxy for the wavelength range 3500 \AA\ to 4500 \AA. Within
this spectrum the wavelength range 3900 \AA\ to 4000 \AA\ has been
masked out. The series of panels show how the reconstruction of this
masked region improves as we iteratively improve the underlying
eigenbasis (with the true spectrum shown as a solid line and the
reconstruction by a dotted line). Figure 8a shows the initial linear
interpolation across the spectral region 3900 \AA\ to 4000 \AA\ from
which the zeroth order eigenbasis, $e^0_i$ is constructed. The rms
dispersion between the true spectrum and the linearly interpolated
reconstruction is 0.019 (which should be treated as the fiducial mark
against which all other reconstruction techniques are applied). The
first order correction, $f_i^1$ (for 3 eigencomponents) is shown in
Figure 8b. At this point the rms dispersion between the true and
reconstructed spectra has already fallen to 0.0095.  For a
reconstruction using 3 eigencomponents the procedure converges rapidly
with a difference between subsequent iterations of $<1\%$ by the fifth
iteration (see Figure 8c). At this stage the reconstruction is stable
with an rms uncertainty of 0.0077 (comparable to the values we derived
when we knew the underlying eigenbasis). We can, therefore, increase
the number of components used in the reconstruction (to improve the
eigenbasis and the interpolation). Figure 8d shows this effect where,
after iterating five times using 3 components, we increase the number
of eigenspectra to five. The dispersion between the reconstructed and
true spectrum falls to 0.0013 with five components.

The number of iterations and components used in the construction is
dependent on the information present within the data.  With the
Bruzual and Charlot models the galaxy spectra can be reconstructed to
an accuracy of less than 1\% using only 5 eigencomponents.  For real spectra
we would expect the number of components to be dependent on the
spectral resolution of the data and the intrinsic wavelength coverage.

\section{Discussion}

We have described a general framework for undertaking spectral
classification of galaxy spectra, accounting for gaps within the data
and different intrinsic rest wavelength coverage. It is expected that
when this technique is applied to galaxy spectra from the next
generation of spectroscopic surveys (such as the SDSS and 2dF) we will
have a mechanism for measuring the spectral evolution of galaxies in
terms of a common classification scheme that is almost independent of
redshift.

Standard techniques for classifying galaxy spectra using eigenspectra
have either ignored the effect of gaps within the data or restricted
their analysis to wavelengths that are common to all galaxies within
their sample. For the next generation of redshift surveys we can
expect the redshift distribution of the galaxies to be broad and, if
we were to restrict our analysis to common wavelengths, the available
spectral range on which we could classify a galaxy to be very
small. Our technique alleviates this problem. It allows a single
eigenbasis to be derived over a very broad intrinsic wavelength range
(from the data themselves) and the classification of the galaxy
spectra to be corrected for the incomplete coverage. The derivation of
the covariance matrix for this correction enables us to determine both
a classification and a measure of the uncertainty on this parameter.

We expect that this classification technique will be equally
applicable to continuum subtracted data (i.e.\ absorption line
spectra) as it is to the spectrophotometric data we analyze here. The
number of eigenspectra required to classify or reconstruct a galaxy
spectrum will be dependent on the quality of the data (i.e.\
resolution and signal-to-noise) and the overall wavelength coverage.
As noted previously the immediate application of this technique will
be to measuring redshifts from local and high-redshift spectroscopic
surveys. The analysis we describe provides an optimal noise
suppression that, when combined with the redshift determination, will
produce a high signal-to-noise representation of noisy and incomplete
data together with an associated error estimate. In a following paper
(Connolly and Szalay 1999) we describe the implementation of our
techniques for constructing spectral energy distributions and
classifying galaxy spectra using redshift survey data.

The application of this technique to astrophysical problems is,
however, substantially more general than providing corrections to
galaxy classification. It can equally well be applied to galaxy
broadband photometry (Csabai et al.\ 1998) or a combination of
spectrophotometry and photometry. In the next few years ground- and
space-based instrumentation will provide broadband photometry and
spectroscopy for large samples of galaxies covering wavelengths from
the ultraviolet through to the far-infrared. The generalization of the
construction of galaxy eigenspectra from noisy data and for spectra
that do not have complete spectral coverage will enable us to
construct a composite eigenbasis, and consequently composite galaxy
spectral energy distributions, that cover a broad spectral range.


\section{Conclusions}

The use of a Principal Component Analysis or KL transform to classify
galaxy spectra is becoming a standard technique in the analysis of
spectroscopic survey data. The current applications of this approach
have assumed that the galaxies we wish to classify and the eigenbasis
we will use for the classification cover the same wavelength
range. For real data this will clearly not be the case. Sky lines that
are masked out of spectra and changes in rest wavelength coverage
due to the redshift of a galaxy will all introduce gaps into the
spectra. Unless we correct for this effect we will introduce
systematic errors into our classification scheme.  In this paper we
derive a generalized form for the KL classification that accounts for
the presence of gaps within galaxy spectra.  We show that applying
this technique to simulated spectra we can determine a robust
classification of galaxy spectra (together with an error) that is
relatively insensitive to the redshift of the galaxy in question.

\acknowledgments
We would like to thank Istvan Csabai for many useful discussions on
the application and interpretation of the gappy KL analysis and the
anonymous referee for comments that helped clarify the technical
discussion.  We acknowledge partial support from NASA grants
AR-06394.01-95A and AR-06337.11-94A (AJC) and an LTSA grant (ASZ). The
SVD analysis was undertaken using the Meschach library of linear
algebra algorithms from Stewart and Leyk.

\clearpage

\begin{figure}
\caption{The first three eigenspectra derived from a simple stellar
population Bruzual and Charlot spectral synthesis model. We construct
a set of galaxy spectra with ages ranging from 0 Gyr to 20 Gyr using
an instantaneous burst model. From these spectra we derive the
eigenbasis using Singular Valued Decomposition. The three panels show
the first three eigenspectra. The first spectrum is the mean of the
galaxy spectra within the sample. The second spectrum correlates with
the star formation and the third spectrum has the spectral features of
an old stellar population.}
\end{figure}

\begin{figure}
\caption{The distribution of eigenvalues for the Bruzual and Charlot
eigenbasis. We show the eigenvalues associated with the first 23
eigenspectra derived from the Bruzual and Charlot spectra.  The first
three eigenspectra contain 99.97\% of the variance (or information)
contained within the Bruzual and Charlot spectra.}
\end{figure}

\begin{figure}
\caption{Reconstructing gappy data using a Karhunen-Lo\'{e}ve
transform. Projecting galaxy spectra onto an eigenbasis and correcting
for the gaps in the spectra we can construct an optimal (in the least
squares sense) interpolation scheme. We show this reconstruction for 3
types of galaxies derived from an instantaneous burst of star
formation Bruzual and Charlot spectral synthesis model. The galaxy
types, moving down from the top panel, are a 0 Gyr, 0.16 Gyr and 20
Gyr spectrum respectively. For the reconstruction the wavelength range
3800 \AA\ to 4000 \AA\ have been masked.  The solid line shows the
spectrum of the complete or ``true'' spectrum the dotted line shows
the reconstructed spectrum. The left hand panel shows the
reconstruction using 3 eigencomponents and the right hand panel the
reconstruction using 5 components.}
\end{figure}

\begin{figure}
\caption{Extrapolating the interpolation scheme to account for
different rest wavelength coverage. For an ensemble of galaxies at a
range of redshift the intrinsic rest wavelength coverage will be
different for each galaxy. Using the eigenbasis we can not just
interpolate over missing data we can also extrapolate to predict the
spectrum even in wavelength regions we do not sample. We demonstrate
this by excluding the data in the wavelength range 7000 \AA\ to 8000
\AA.  As in Figure 3 we show this extrapolation for three types of
galaxy, with ages 0 Gyr, 0.16 Gyr and 20 Gyr respectively. The solid
line shows the true spectrum and the dotted line the reconstructed
spectrum. The left panel is for a reconstruction using 3
eigencomponents and the right panel for 5 components.}
\end{figure}

\begin{figure}
\caption{Projecting galaxy spectra onto their eigenbasis provides a
simple and physical means of classifying galaxies. If we do not
account for the gaps within the data we introduce both random and
systematic errors into this classification scheme. In Figure 5a we
show the intrinsic correlation of galaxy type with the first two
expansion coefficients ($a_1$ and $a_2$) as a solid line. The effect
of excluding 10\% of the data (in ten intervals of 45 \AA\ randomly
positioned within the spectrum) is shown by the crosses. Correcting
for the gappy nature of the data we can reconstruct the original
classification coefficients (solid ellipses). Figure 5b shows the
effect of redshift on the classification. By randomly removing up to
20\% of the spectrum at the long wavelength end of the spectrum we
introduce substantial systematic errors into the classification
scheme.  As with Figure 5a the solid ellipses show the coefficients
once they have been corrected for the gaps within the spectra. The
size of the ellipses in the plot represents the three sigma error in
the corrected coefficients.}
\end{figure}

\begin{figure}
\caption{An optimal filtering of noisy data. By projecting a noisy
spectrum onto it s eigenbasis we can reconstruct an optimal filtered
spectrum (in the least squares sense). The top figure shows a 2 Gyr
spectrum with a signal-to-noise of approximately 5 per pixel. The
lower figure shows the ``true'' spectrum (solid line) together with
the filtered spectrum (dotted line). The filtering of the noisy
spectrum is undertaken using 3 eigencomponents. }
\end{figure}

\begin{figure}
\caption{An optimal filtering of noisy data in the presence of gappy
data. As in Figure 6 we reconstruct a filter galaxy spectrum using 3
eigencomponents. The upper figure shows the spectrum with a
signal-to-noise of approximately 5. The spectral region 3800 \AA\ to
4000 \AA\ has been masked out of this spectrum. The lower figure shows
the reconstructed spectrum and the interpolation across the region of
missing data (dotted line). As before the interpolation was undertaken
with 3 eigencomponents.}
\end{figure}

\begin{figure}
\caption{Building spectral energy distributions from gappy data. Given
a set of galaxy spectra with missing data (ten gaps of length 100 \AA\
were randomly placed into each spectrum) we can still recover the
underlying eigenbasis. We undertake this in an iterative fashion (see
text for details).  Figure 8a shows a 2 Gyr old spectrum (solid line)
and the reconstructed spectrum, across the spectral region 3900 \AA\
to 4000 \AA\ using linearly interpolation (dotted line). Figure 8b
shows the first reconstruction using 3 eigencomponents and Figure 8c
the reconstruction after 5 iterations. Figure 8d shows the effect of
adding additional components (7) to the reconstruction after the fifth
iteration.}
\end{figure}

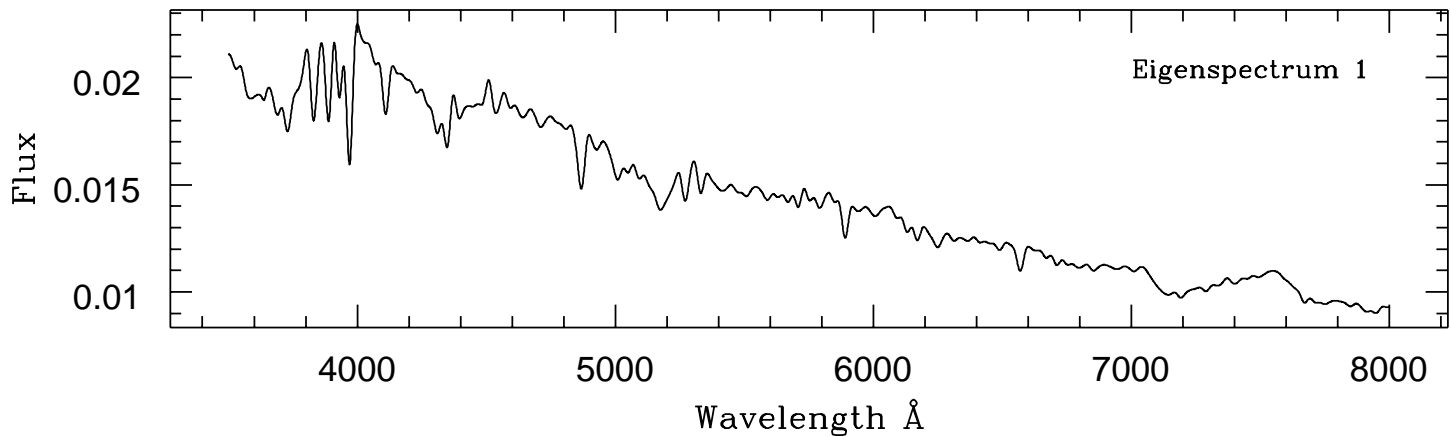
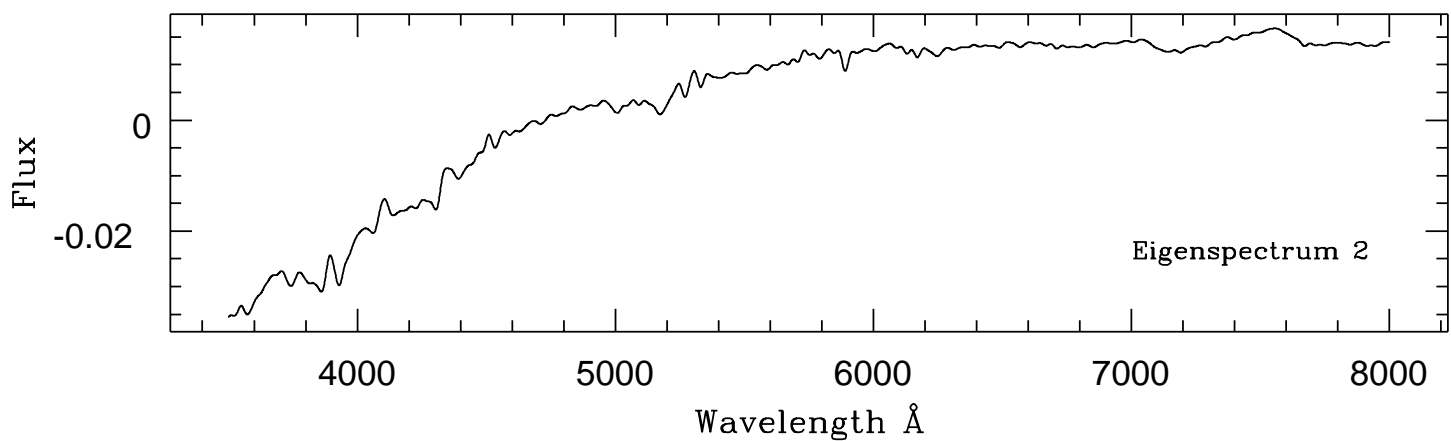
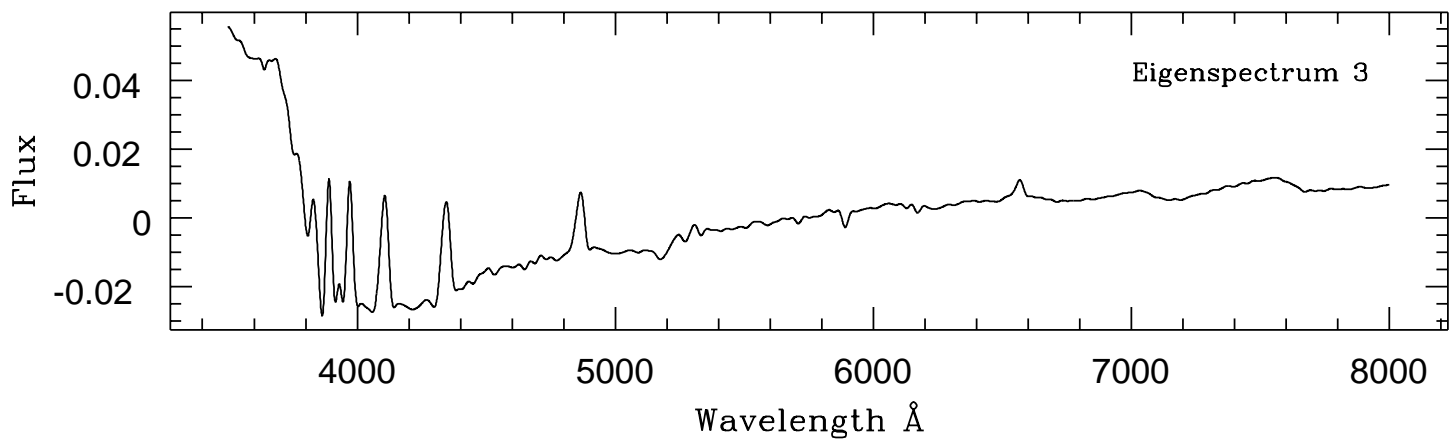
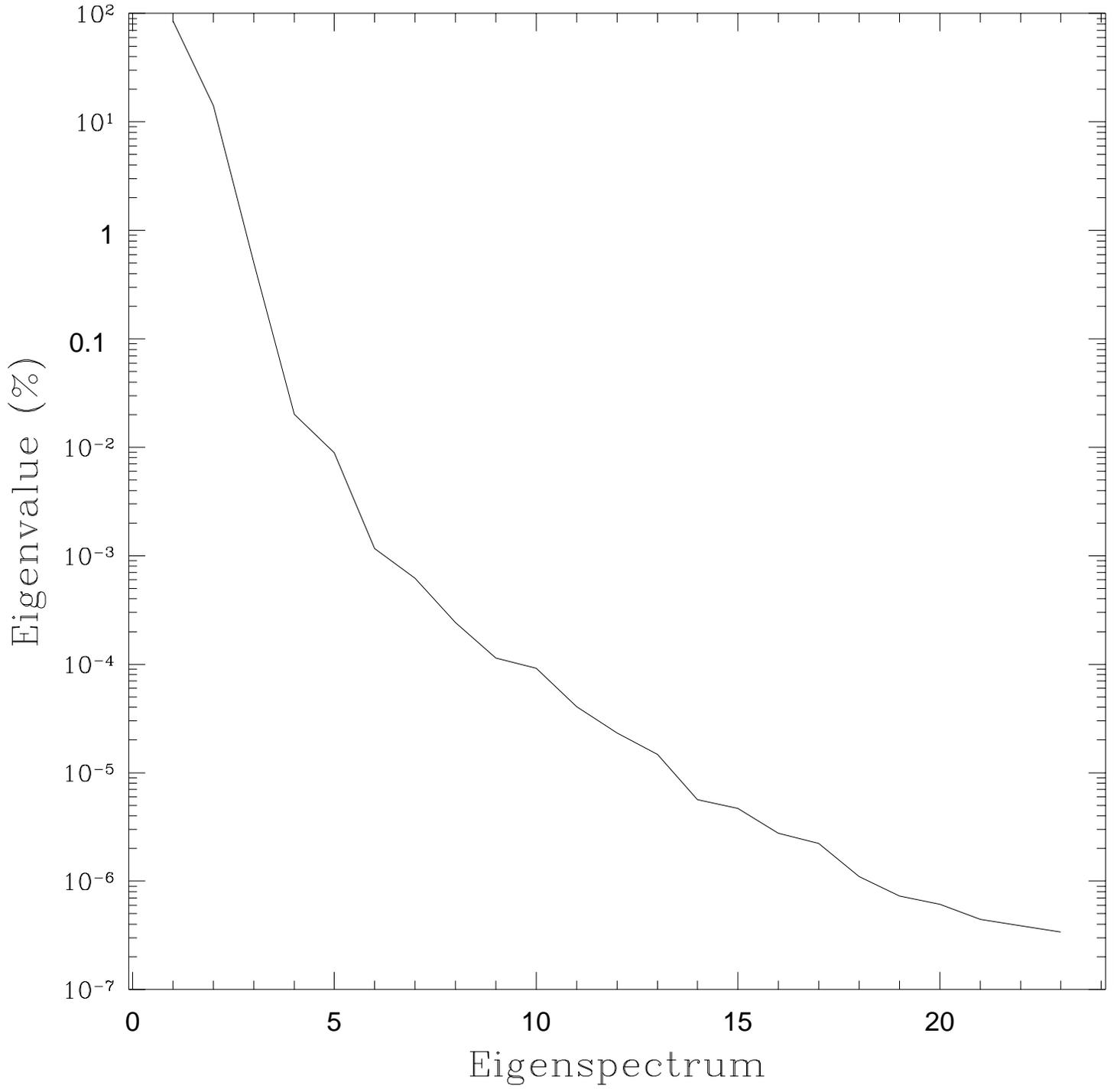
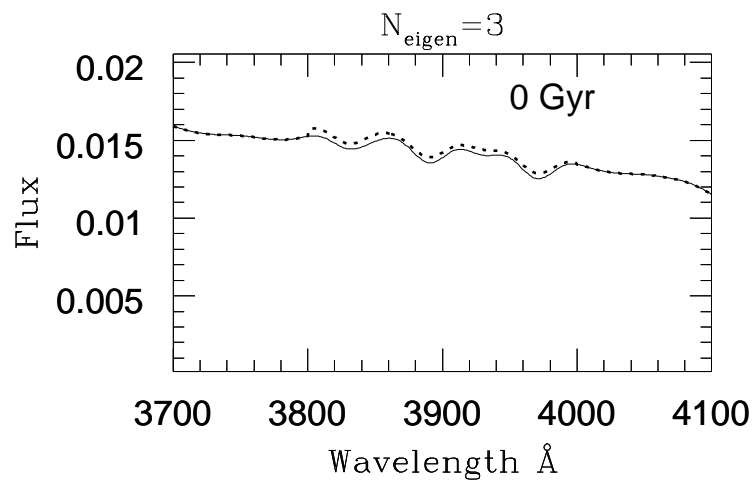
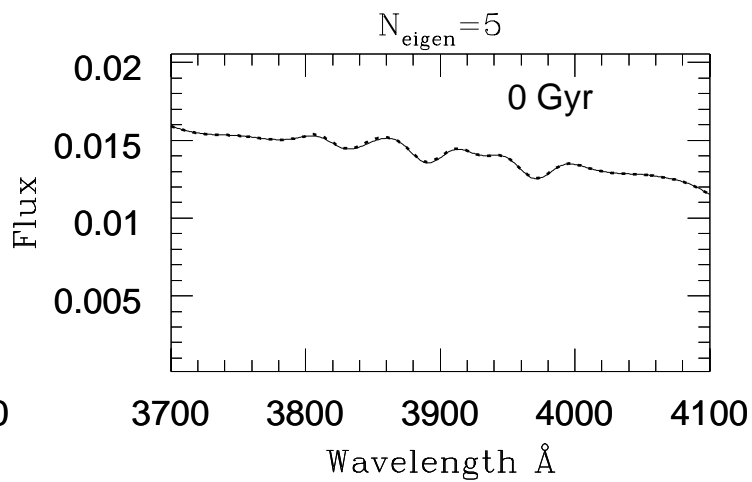
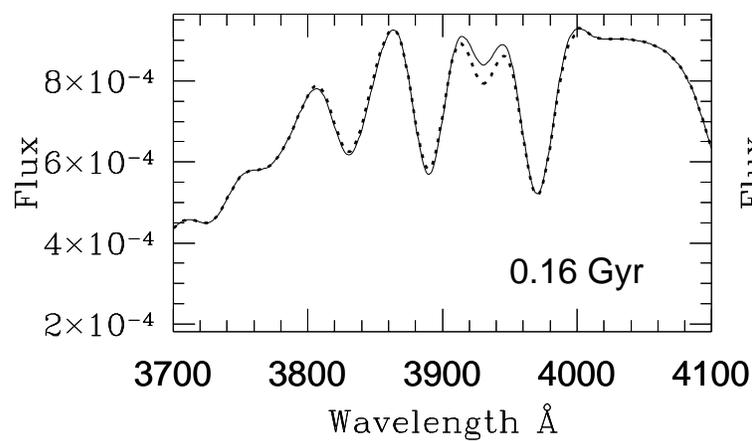
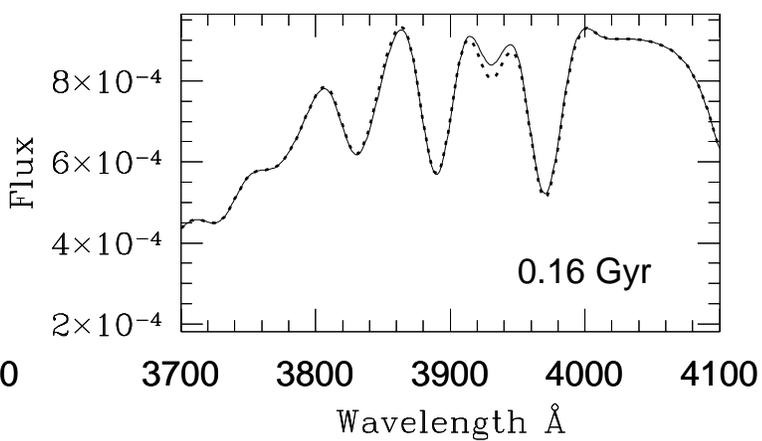
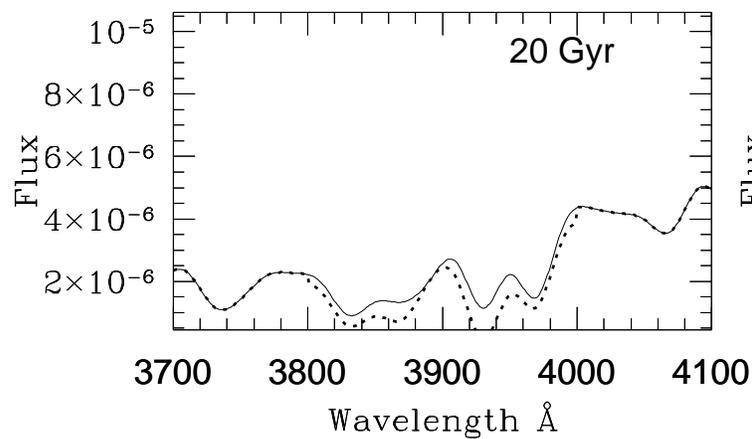
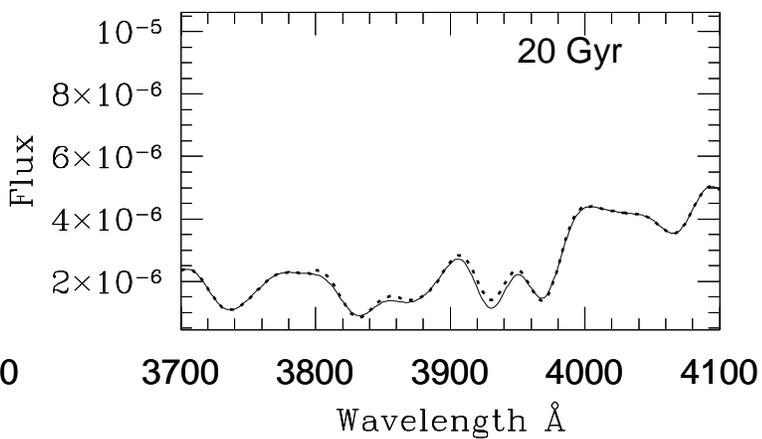
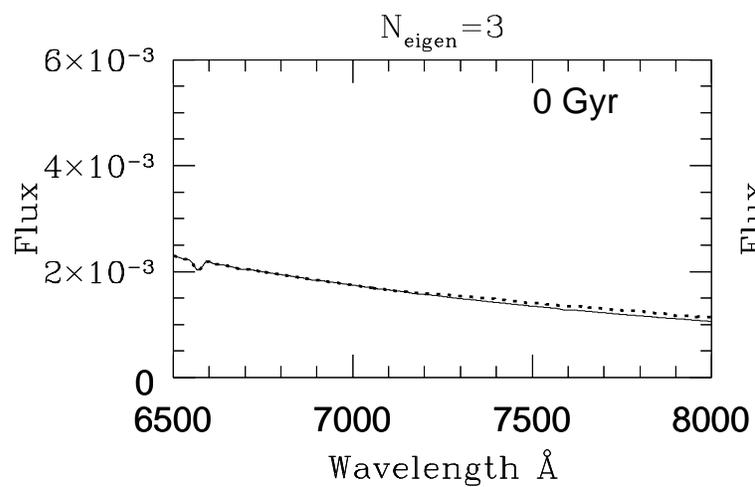
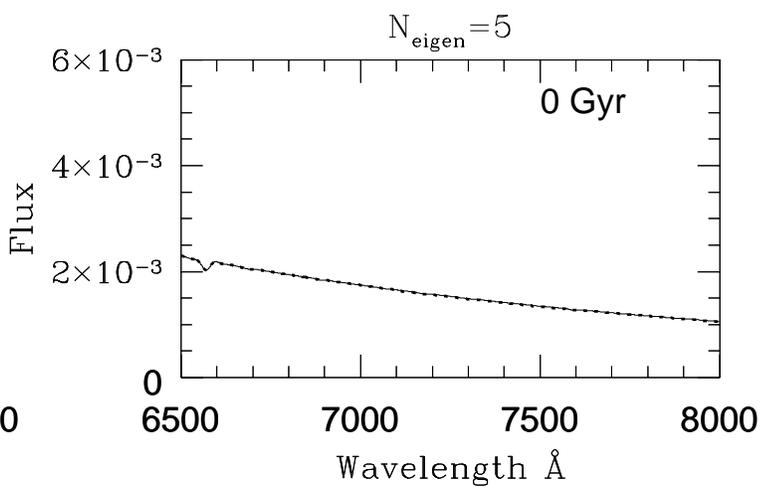
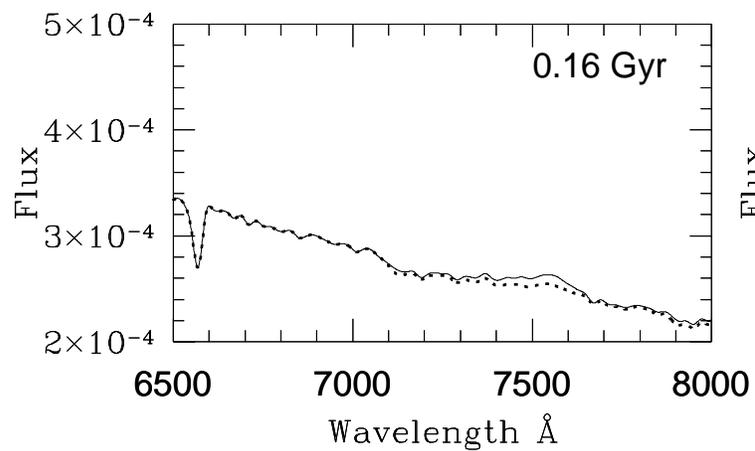
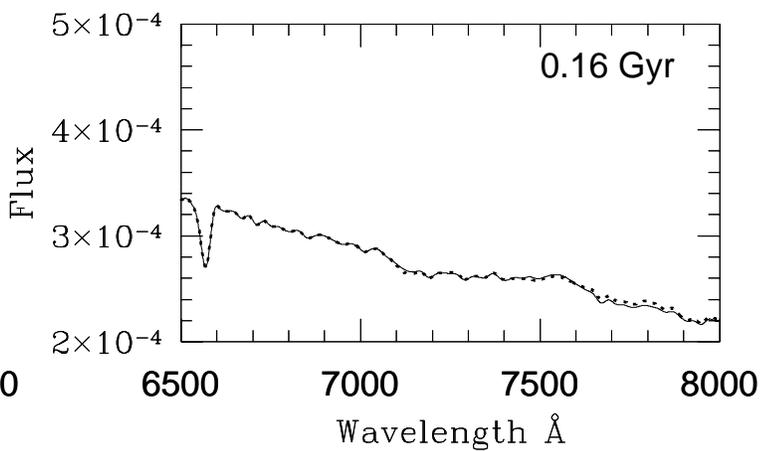
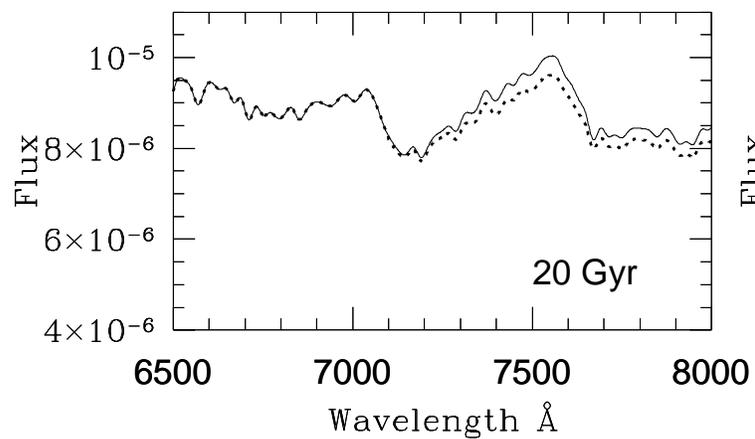
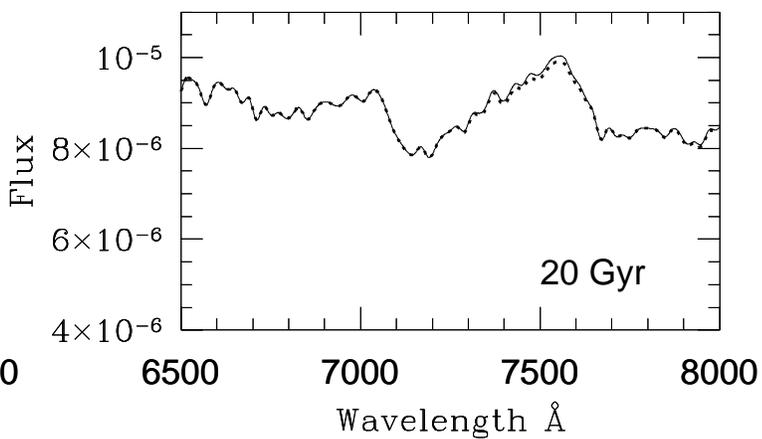
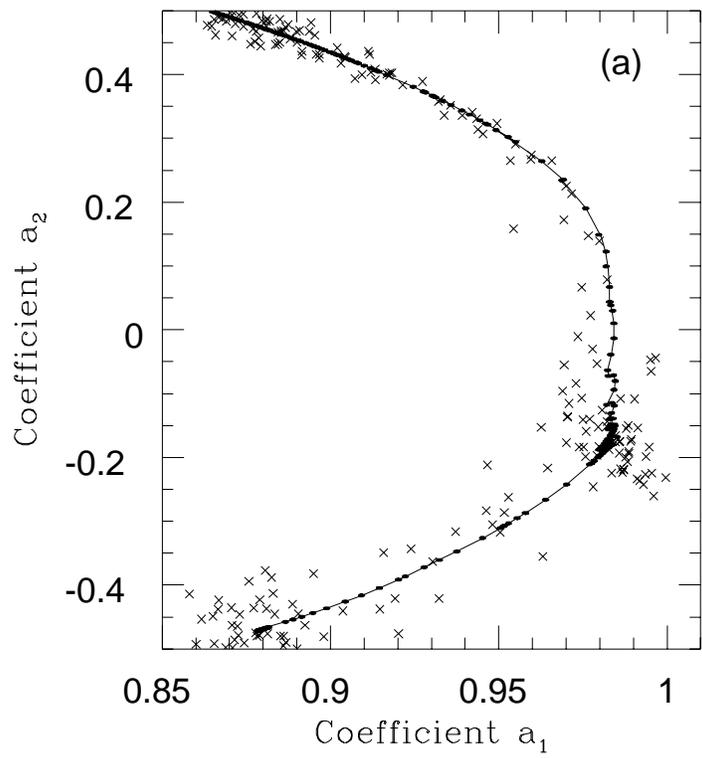
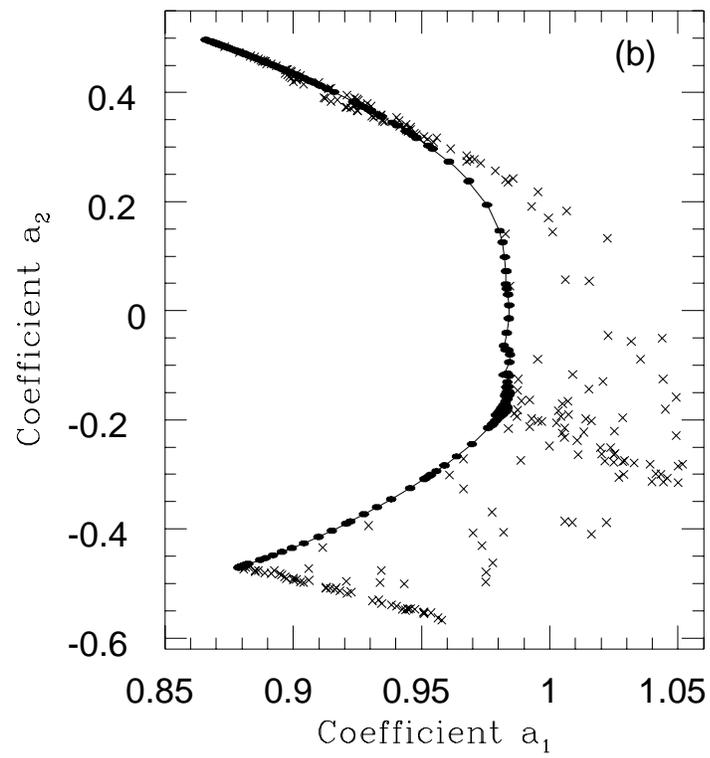
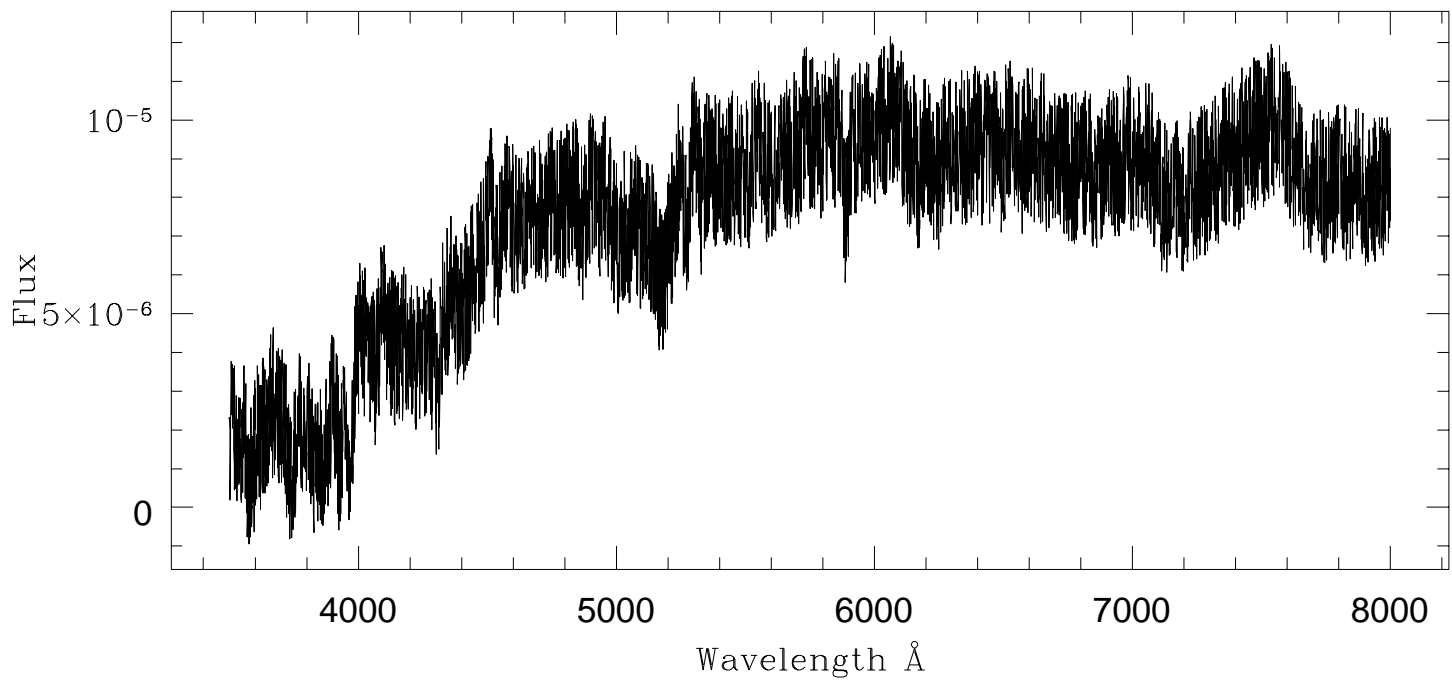
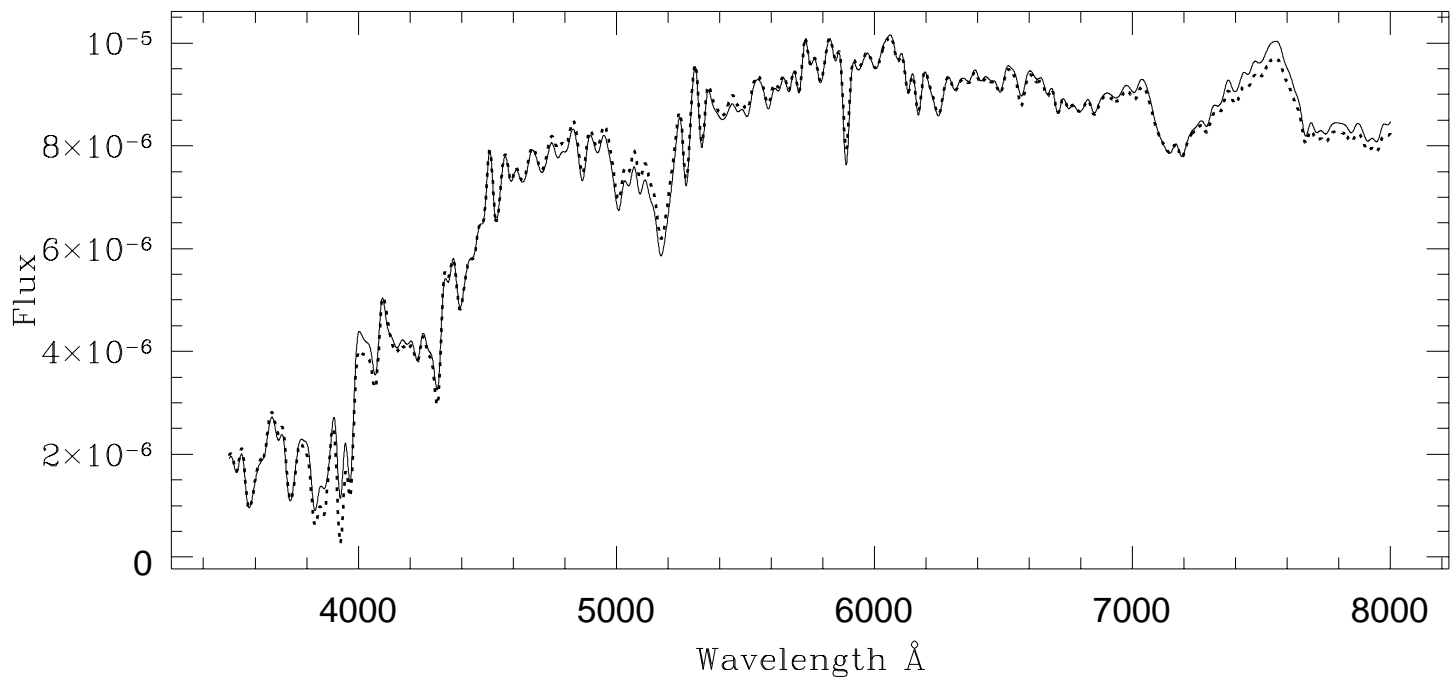
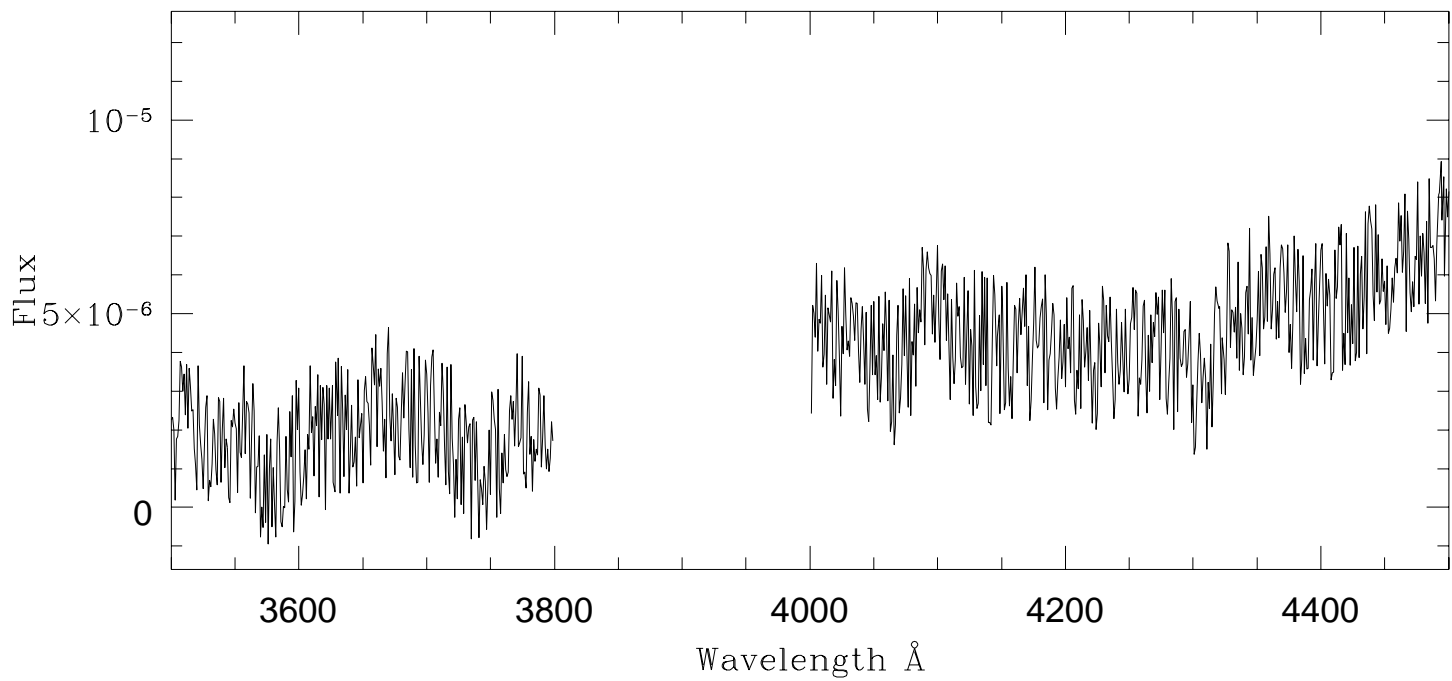
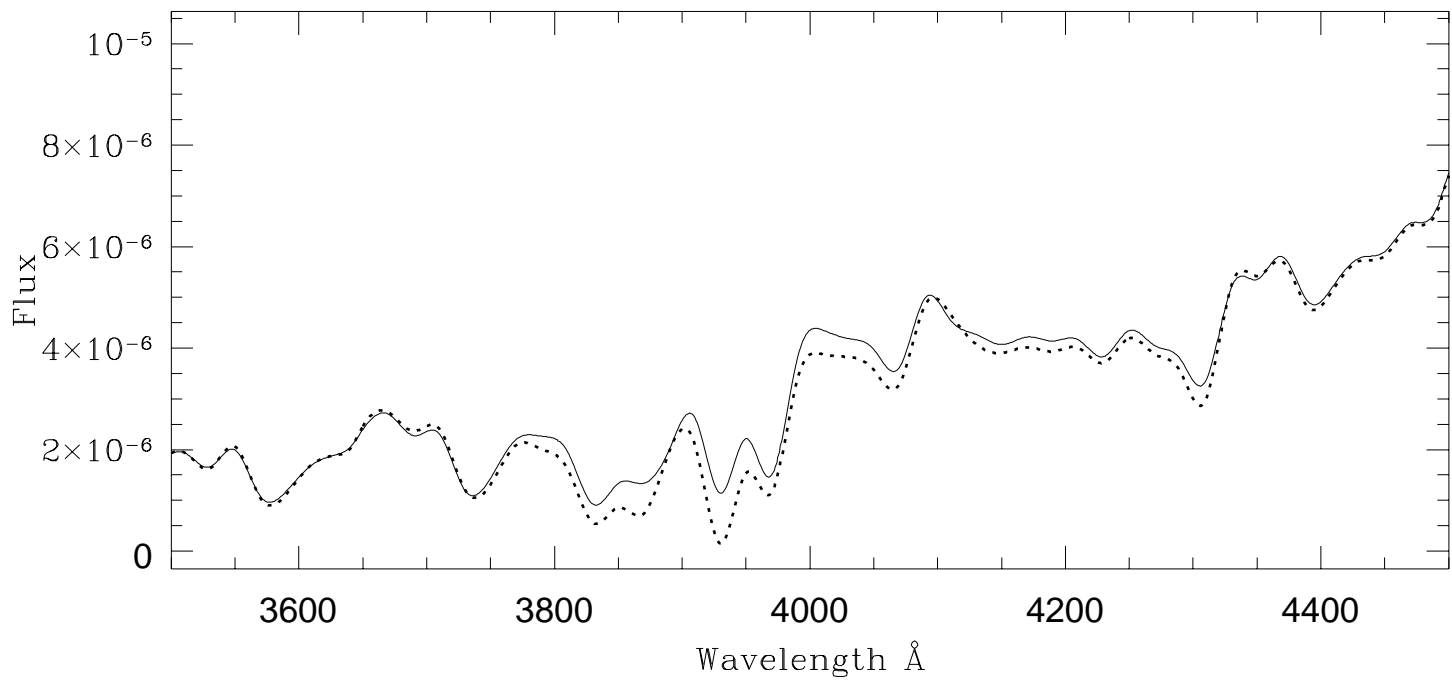
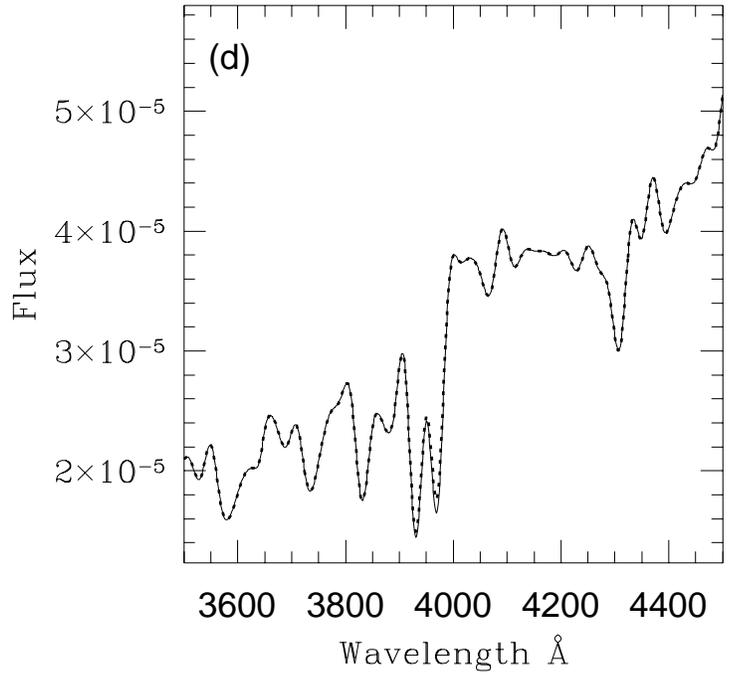
\end{document}